\documentclass[11pt,superscriptaddress,reprint, preprintnumbers,nofootinbib,prd,twocolumn]{revtex4-2}
\pdfoutput=1 
\usepackage[dvipsnames,table,xcdraw]{xcolor}
\usepackage{amsmath,amssymb,slashed,mathtools}
\usepackage{graphicx,multirow}

\usepackage{mathtools,halloweenmath}
\usepackage{multirow}
\usepackage{soul}

\usepackage[table]{xcolor}
\newcommand{\cmark}{\ding{51}}%
\newcommand{\xmark}{\ding{55}}%
\newcommand{\qn}[3]{({\bf #1},{\bf #2}, #3)}%

\usepackage{tikz}
\usetikzlibrary{decorations.markings,decorations.pathmorphing}
\usepackage{pifont}

\usepackage[toc]{appendix} 
\usepackage{verbatim}
\usepackage[colorlinks,citecolor=blue]{hyperref}

\begin{document}

\title{
Decoupling Neutrino Magnetic Moment from Mass with \texorpdfstring{$SU(2)_L$}{SU2L} Invariance}

\author{Julie Pag\`{e}s}
\email{ jcpages@ucsd.edu}
\affiliation{Physics Department, University of California San Diego,
9500 Gilman Drive, La Jolla, CA 92093, USA}

\author{Anil Thapa}
\email{a.thapa@colostate.edu}
\affiliation{Physics Department, Colorado State University, Fort Collins, CO 80523, USA}

\hypersetup{
pdftitle={Magnetic Moment},   
pdfauthor={Anil Thapa, Julie Pages}
}

\begin{abstract}
Standard Model extensions that yield observable neutrino magnetic moments typically also induce large neutrino masses, incompatible with experimental limits. 
This tension motivates the search for mechanisms that naturally decouple magnetic moments from mass generation without requiring fine-tuning.
In this letter, we propose a novel mechanism for generating Dirac and Majorana neutrino magnetic moment, in which the associated mass contribution is forbidden by $SU(2)_L$ invariance.
By carefully selecting the $SU(2)_L$ representations connecting the neutrino to the loop diagram, we ensure that only the effective dipole operator involving the non-Abelian part of the photon --- the neutral $SU(2)_L$ gauge boson --- is generated. Crucially, the corresponding mass diagram, obtained by removing the external gauge boson leg, vanishes.
We provide explicit UV completions that implement this mechanism and yield neutrino magnetic moments within the sensitivity of current and future experiments. 
\end{abstract}

\maketitle

\section{Introduction}
\label{intro}
The observational evidence of neutrino oscillations~\cite{ParticleDataGroup:2024cfk} indicates that the Standard Model (SM) of particle physics is incomplete, suggesting the need for physics beyond the Standard Model (BSM). In such BSM theories, neutrinos may acquire electromagnetic properties -- such as electric charge, charge radius, and magnetic moment -- through quantum corrections. These properties thus offer a sensitive and complementary avenue for probing BSM physics (for a recent review, see Ref.~\cite{Giunti:2024gec}). Among them, the neutrino magnetic moment is particularly interesting due to its connections with the origin of neutrino masses and CP properties~\cite{Schechter:1981hw,Nieves:1981zt,Kayser:1982br,Kayser:1984ge}. It may also play an important role in distinguishing whether neutrinos are Dirac or Majorana particles~\cite{deGouvea:2022znk}. Furthermore, neutrinos with large magnetic moments can lead to observable consequences in terrestrial experiments and astrophysical environments, such as in neutrino scattering~\cite{Beda:2012zz,Borexino:2017fbd,CONUS:2022qbb}, dark matter direct detection experiments~\cite{XENON:2020rca,XENON:2022ltv,LZ:2022lsv}, collider experiments~\cite{Boyarkin:2014hva,Frigerio:2024jlh,Brdar:2025iua}, astrophysical neutrino signals~\cite{Ahriche:2003wt,Akhmedov:2003fu,Jana:2022tsa,Akhmedov:2022txm,Kopp:2022cug,Ando:2002sk,Brdar:2023cub,Jana:2023ufy}, stellar cooling rates~\cite{Viaux:2013lha,Viaux:2013hca,Capozzi:2020cbu}, and cosmological imprints~\cite{Vassh:2015yza,Li:2022dkc,Carenza:2022ngg,Grohs:2023xwa}.

One of the major theoretical challenges is to obtain a large magnetic moment while remaining consistent with the observed small masses of  neutrinos, $m_\nu < 0.45 $~eV \cite{KATRIN:2024cdt}. The effective vertex that generates the dipole operator in any effective field theory (EFT) gives rise to neutrino magnetic moments $\mu_\nu$, and in general also induces neutrino masses $m_\nu$ when the external photon leg is removed.
This results in the following estimate
 \begin{align}
     m_\nu \sim \frac{\mu_\nu}{\mu_B} \frac{\Lambda^2}{2 m_e} \, , 
     \label{eq:numass_NMM}
 \end{align}
 where $\Lambda$ is the scale of the heavy particle(s) inside the loop, $\mu_B\equiv e / 2 m_e$ is the Bohr magneton, $e$ is the electric charge, and $m_e$ is the electron mass.
Since at least one new particle running inside the loop must be charged, and thus above $100$ GeV to avoid direct search limits, this leads to the naturalness bound $\mu_\nu~\lesssim~10^{-19} \mu_B\ \left(\frac{\delta m_\nu}{0.1\text{eV}}\right)\left(\frac{\Lambda}{\text{TeV}}\right)^{-2}$ \cite{Fujikawa:1980yx,Pal:1981rm,Shrock:1982sc,Dvornikov:2003js,Dvornikov:2004sj}. In contrast, the current experimental limit, $\mu_\nu^{\rm exp} \lesssim 10^{-11} \mu_B$ \cite{Cowan:1954pq,Vidyakin:1992nf,Derbin:1993wy,MUNU:2005xnz,TEXONO:2009knm,Allen:1992qe,Beda:2012zz,Borexino:2017fbd,CONUS:2022qbb}, is several orders of magnitude higher. \footnote{See Ref.~\cite{Canas:2015yoa} for a review on global fits including all experimental limits.} Additionally, the energy loss of the red giant branch in the globular cluster provides the most stringent limit of $\mu_\nu^{\rm astro} < 4.5 \times 10^{-12} \mu_B$ \cite{Viaux:2013lha}. Consequently, generating a neutrino magnetic moment near the experimental sensitivity requires severe fine-tuning of one part in $10^7$.  

To consistently generate a large neutrino magnetic moment, it is necessary to identify mechanisms that break the correlation given in Eq.~\eqref{eq:numass_NMM}. Two prominent mechanisms have been proposed in the literature:
(a) Spin suppression in the BFZ-like model results in magnetic moment generation at two loops via the $\gamma h^+ W^-$ effective vertex \cite{Barr:1990um,Barr:1990dm,Babu:1992vq}. Without the photon line, the effective vertex would describe a spin-0 to spin-1 transition, forbidding transversely polarized $W$ bosons. Consequently, the corresponding mass diagram is suppressed by an additional factor $m_\ell^2/m_W^2$ with respect to the magnetic moment diagram. (b) Voloshin-type symmetry applied to Majorana neutrinos that employs an $SU(2)_H$ horizontal flavor symmetry to generate a large magnetic moment while naturally suppressing neutrino masses \cite{Babu:1990wv,Leurer:1989hx}. This exploits the fact that the Majorana dipole operator is antisymmetric in flavor space whereas the Majorana mass term is symmetric.  Note that the original proposal by Voloshin~\cite{Voloshin:1987qy,Barbieri:1988fh} for Dirac neutrinos imposed an $SU(2)_\nu$ symmetry with $(\nu_R^C, \nu_L)^T$ transforming as doublet. Achieving a neutrino magnetic moment of $\mu_\nu \sim 10^{-12} \mu_B$ in this case requires fine-tuning at the level of four orders in magnitude \cite{Lindner:2017uvt}.
  
In this letter, we propose a new mechanism based on the existing $SU(2)_L$ gauge symmetry of the SM such that the effective dipole operator exists only for the $SU(2)_L$ gauge bosons $W_\mu$ (triplet of $SU(2)_L$) but not for the $U(1)_Y$ hypercharge gauge boson $B_\mu$ (singlet of $SU(2)_L$), since the photon field $A_\mu$ is a linear combination of the neutral fields $W_\mu^3$ and $B_\mu$. The contribution to the neutrino mass without any gauge boson attached is group theoretically identical to the $B_\mu$ graph and therefore also vanishes due to gauge invariance, thus evading the correlation of Eq.~\eqref{eq:numass_NMM}. We begin by providing a comprehensive analysis of the effective vertices involving only $W_\mu$ bosons. We then open up such effective vertices by introducing new heavy degrees of freedom and demonstrate that several UV complete models allow for potentially observable neutrino magnetic moments. We emphasize that our aim is not to eliminate the RGE-induced contributions~\cite{Bell:2005kz,Davidson:2005cs,Bell:2006wi} that are universal and unavoidable but to construct a mechanism within the SMEFT framework and then UV-complete it, where the matching at the UV scale yields a sizable $W$-dipole contribution while maintaining a vanishing mass contribution.

\section{Mechanism to suppress neutrino mass}\label{sec:suppressmass}
\begin{figure}[t!]
    \begin{tikzpicture}
    \tikzset{
                fermion/.style={draw=black, line width=1.0pt, postaction={decorate},
                },
                 fermion_no_arrow/.style={draw=red, line width=1.0pt},
                scalar/.style={draw=black, dashed, line width=1.0pt},
                photon/.style={decorate, decoration={snake, amplitude=2pt, segment length=6pt}, draw=black, line width=1.0pt},
                scalar/.style={draw=black, dashed, line width=1.0pt}
            }
    \draw[fermion] (-1.5,0) 
    -- (-0.4,0) node[midway, above] {$R_1$}  ;
    \draw[fill=gray!30] (0,0) circle (0.4);
    \draw[fermion] (0.4,0)  -- (1.5,0) node[midway, above] {$R_2$} ;
    \draw[photon] (0,0.5) -- (0,1.1) node[midway, left] {3} node[above] {$W_\mu$};
    \draw[scalar] (-0.3,-0.3) -- (-1,-1) node[midway, above left] {2} node[below left] {$H$};
    \draw[scalar] (0.3,-0.3) -- (1,-1) node[midway, above right] {2} node[below right] {$H$};
    \draw[dotted] (0.5,-.8) arc[start angle=315, end angle=225, radius=0.8] node[ midway, below] {$n$};
    \node at (0,-1.8) {(a)};
    \end{tikzpicture}
      ~~~~~~~~~
     \begin{tikzpicture}
    \tikzset{
                fermion/.style={draw=black, line width=1.0pt, postaction={decorate}, 
                },
                 fermion_no_arrow/.style={draw=red, line width=1.0pt},
                scalar/.style={draw=black, dashed, line width=1.0pt},
                photon/.style={decorate, decoration={snake, amplitude=2pt, segment length=6pt}, draw=black, line width=1.0pt},
                scalar/.style={draw=black, dashed, line width=1.0pt}
            }
    \draw[fermion] (-1.5,0) node[left, below] {$R_{1,2}$} 
    -- (-0.2,0)  ;
    \draw[fill=gray!30] (0,0) circle (0.2);
    \draw[fermion] (0.2,0)  -- (1.5,0) node[right, below] {$L$} node[midway, below] {$2$}  ;
    \draw[scalar] (-0.15,0.15) -- (-1,1) node[midway,  left=2pt] {2} node[above left] {$H$};
    \draw[scalar] (0.15,0.15) -- (1,1) node[midway, right=2pt] {2} node[above right] {$H$};
    \draw[dotted] (-0.6,.8) arc[start angle=135, end angle=45, radius=0.8] node[ midway, above] {$k$};
    \node at (0,-1.8) {(b)};
    \end{tikzpicture}
\caption{(a) Effective vertex with $n$ Higgses, two fermions and a $W_\mu$ gauge boson  with the corresponding $SU(2)_L$ representation $({\bf 2}, {\bf R_1}, {\bf R_2}, {\bf 3})$. (b) Effective vertex to connect $({\bf R_{1}}$ or ${\bf R_{2}})$ to the SM lepton doublet with $k=R_{1,2}-2$ Higgses. A similar diagram exists with the right-handed singlet neutrino $\nu_R$ and $R_{1,2}$ by taking $k=R_{1,2}-1$ Higgses.} \label{fig:mechanism}
\end{figure}
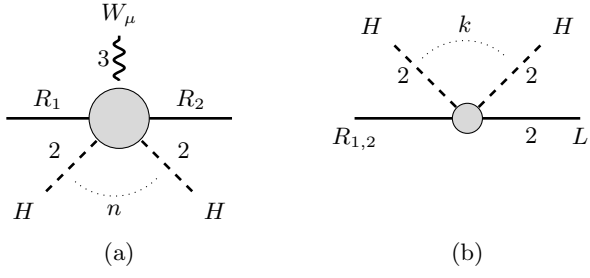 

In the $\nu$SMEFT (the SMEFT augmented with SM singlet right-handed neutrinos $\nu_R$) \cite{delAguila:2008ir,Liao:2016qyd}, the Dirac neutrino magnetic moment arises from dimension-6 operators
\begin{align}
     C_{\nu B} g_1 \bar{L} \widetilde{H} \sigma^{\mu \nu} \nu_R B_{\mu\nu} + C_{\nu W}  g_2 \bar{L} \tau^a \widetilde{H} \sigma^{\mu \nu} \nu_R W_{\mu\nu}^a + \text{h.c.}\, , 
    \label{eq:moment_operator_Dirac}
\end{align}
where $L$ is the SM lepton doublet, $\tilde{H} \equiv i \tau_2 H^*$ is the conjugate Higgs doublet,
$B_{\mu\nu} = \partial_\mu B_\nu - \partial_\nu B_\mu$ and $W^a_{\mu\nu} = \partial_\mu W_\nu^a - \partial_\nu W_\mu^a - g_2 \epsilon^{abc} W_\mu^b W_\nu^c$, $a=1,2,3$, are the field strength tensors of $U(1)_Y$ and $SU(2)_L$, respectively, with associated couplings $g_1$ and $g_2$, and  $\tau^a$ are the Pauli matrices. The  effective dimension-5 Dirac neutrino magnetic moment operator below the electroweak scale is $\mu_\nu^D/2\ \bar{\nu}_L \sigma^{\mu\nu}\nu_R F_{\mu \nu} + \text{h.c.}$, 
where \( F_{\mu\nu} = \partial _\mu A_\nu - \partial_\nu A_\mu \) is the photon field strength tensor. 
The dimension-4 and dimension-6 operators contributing to the Dirac neutrino masses are
\begin{align}
    Y_{\rm Dirac} \,\bar{L} \widetilde H \nu_R\, + C_{\nu H} (\bar{L} \widetilde H \nu_R) (H^\dagger H) .
    \label{eq:mass_operator_Dirac}
\end{align}
In contrast to the Dirac case, the magnetic moments for Majorana neutrinos originate from lepton number violating dimension-7 operators in the SMEFT:
\begin{align}
     & C_{LHB}\ g_1 (\bar{L}^c \epsilon H) \sigma^{\mu\nu} (H^T \epsilon L) B_{\mu\nu} \notag \\
    &+ C_{LHW}\ g_2 (\bar{L}^c \epsilon H) \sigma^{\mu\nu} (H^T \epsilon \tau^i L) W_{\mu\nu}^i \, ,
    \label{eq:moment_operator_Majorana}
\end{align}
where $\epsilon =  i \tau_2$ is the antisymmetric $SU(2)_L$ tensor and $\bar{L}^c \equiv L^T C$ where $C$ denotes the charge conjugation matrix. After electroweak symmetry breaking, a linear combination of these two operators leads to the Majorana neutrino magnetic moment operator, defined as $\mu_\nu^M/4\ (\bar{\nu}_L^c \sigma^{\mu\nu}  \nu_L) F_{\mu\nu}$ with an additional factor $1/2$ to avoid double counting. The dimension-5 and dimension-7 Weinberg operators that generate Majorana neutrino masses are
\begin{align}
    C_{LH}^5 (\bar{L}^c \widetilde H^*)(\widetilde H^\dagger  L)\, + C_{LH}^7 (\bar{L}^c \widetilde H^*)(\widetilde H^\dagger  L) (H^\dagger H) .
    \label{eq:mass_operator_Majorana}
\end{align}
The Majorana neutrino transition magnetic moment $[\mu_\nu^M]_{\alpha \beta}$ is antisymmetric\footnote{The Wilson coefficient $C_{LHB}$ is antisymmetric under the flavor indices, while $C_{LHW}$ is neither symmetric nor antisymmetric. However, it can be decomposed as $[O^\pm_{LHW}]_{ab} = 1/2 ([O_{LHW}]_{ab} \pm [O_{LHW}]_{ba} )$.  Only the antisymmetric part contributes to the magnetic moment after electroweak symmetry breaking. A fully antisymmetric flavor structure can be constructed as $(\bar{L}^c \epsilon \tau^a \sigma_{\mu\nu}L) (H \epsilon \tau^b H) W_d^{\mu\nu} \varepsilon_{abd}$ \cite{Davidson:2005cs,Bell:2006wi}.
}  in its flavor indices $\{\alpha, \beta \}$, whereas the mass terms are flavor symmetric. 

 \begin{figure*}[!t]
    \centering
    \includegraphics[scale=0.2]{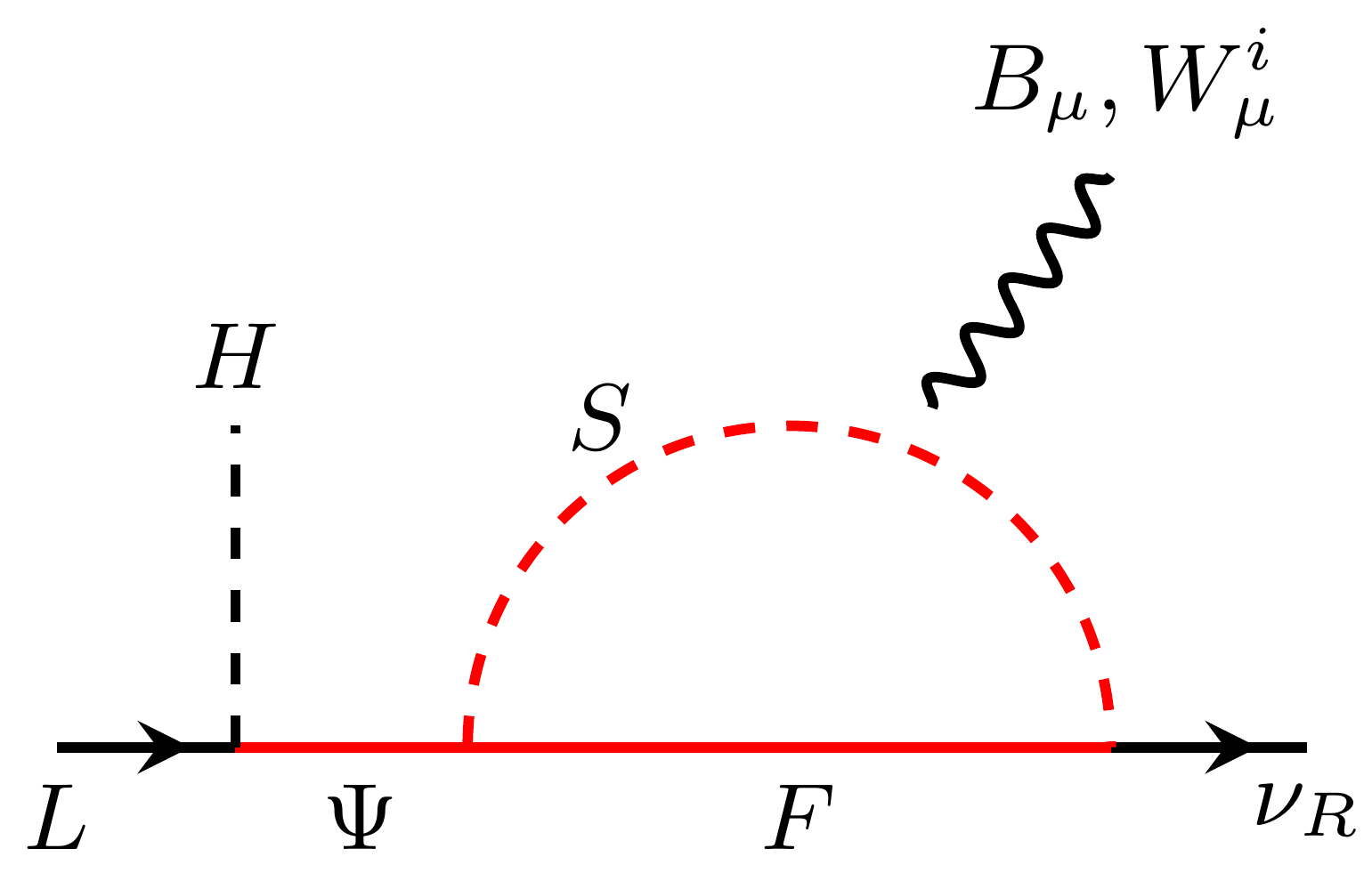} \hspace{15mm}
    \includegraphics[scale=0.23]{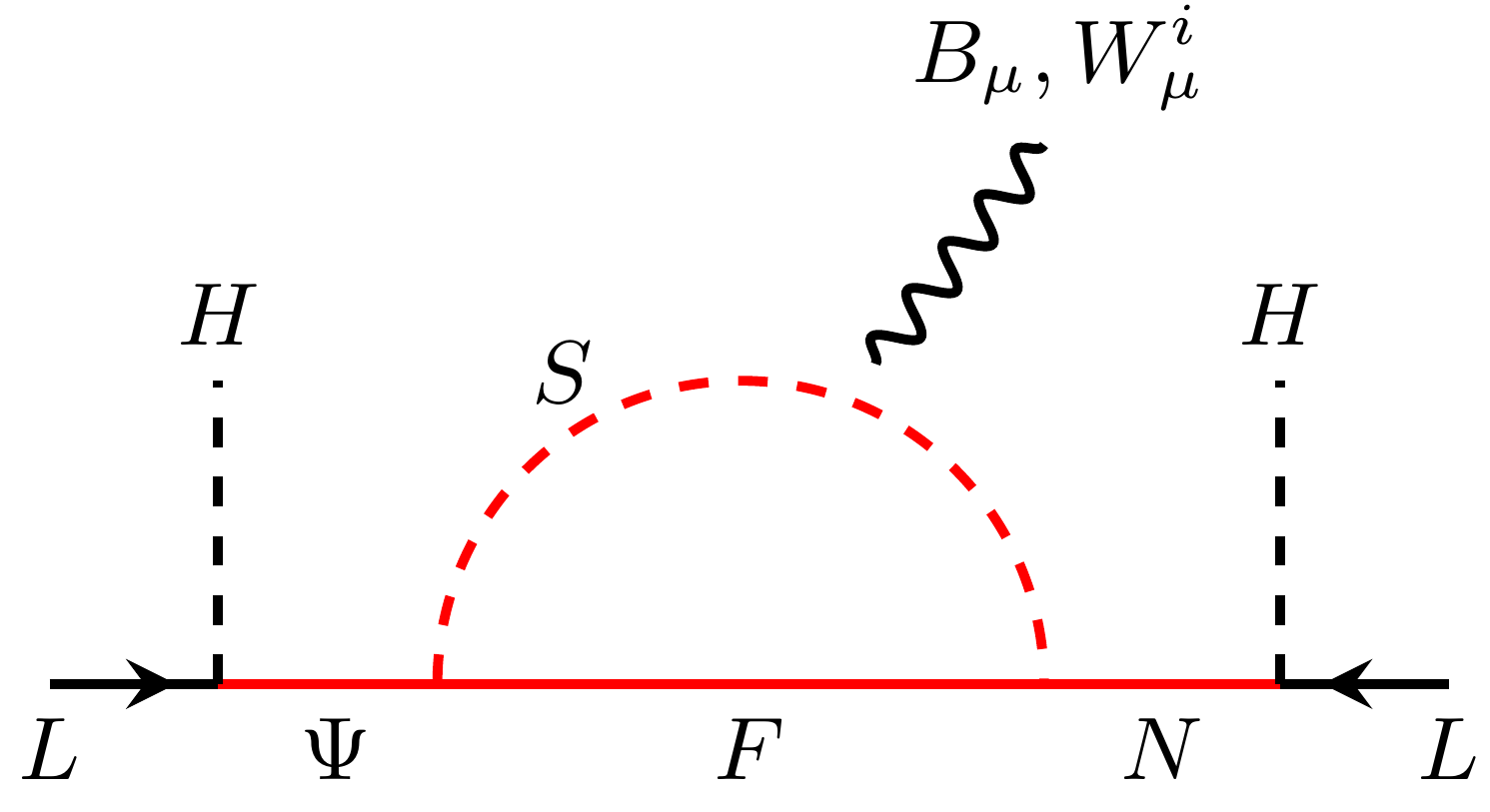} \\
    (a)~~~~~~~~~~~~~~~~~~~~~~~~~~~~~~~~~~~~~~~~~~~~~~~~~~~~~~~~~~~~~~~~~~\,(b)
    \caption{One-loop topology for (a) Dirac and (b) Majorana magnetic moment that leads to vanishing neutrino masses. Black (red) lines represent SM (new) particles. $\Psi$ is a triplet of $SU(2)_L$. The gauge bosons ($B_\mu, W_\mu^i$) can be attached to either scalar line $S$ or fermion line $F$. }
    \label{fig:topology}
\end{figure*} 

Since both $L$ and $H$ transform as doublets under $SU(2)_L$, they can combine to form gauge-invariant operators involving both the singlet gauge boson $B_\mu$ and the triplet gauge boson $W_\mu$, i.e., $\bf 2 \otimes 2 = 3 \oplus 1$, as can be seen from Eqs.~\eqref{eq:moment_operator_Dirac} and~\eqref{eq:moment_operator_Majorana}. The corresponding mass operators --- obtained by removing the field strengths --- are also gauge invariant and lead to the usual correlation between mass and magnetic moments written in Eq.~\eqref{eq:numass_NMM}. The key idea of the proposed mechanism is to generate the dipole operators Eqs.~\eqref{eq:moment_operator_Dirac} and~\eqref{eq:moment_operator_Majorana} via new mediator fermion fields that transform under different (\textit{i.e.} not doublet) representation of $SU(2)_L$, labeled $\bf R_{1,2}$. We then search for effective vertices that contain an invariant in the product $\bf R_1 \otimes R_2 \otimes 3 \,\otimes 2^n$, but not in the product $\bf R_1 \otimes R_2 \,\otimes 2^n$. Here, $\bf 3$ and $\bf 2$ correspond to the $SU(2)_L$ representations of the gauge boson $W_\mu^i$ and the Higgs $H$, respectively. We allow for multiple number of Higgses. A representative diagram illustrating the mechanism is shown in Fig.~\ref{fig:mechanism} and the condition for realizing the magnetic moment through the $W_\mu$ diagram, together with a vanishing mass contribution, is given by
\begin{equation}
    |R_1-R_2|-n=2\,,
\end{equation}
where $n$ is the number of Higgses attached to the effective vertex.
The new fermion(s) acts as a bridge between the effective vertex Fig.~\ref{fig:mechanism}\,(a) and the SM fermions. In practice, to connect the fermion with representation ${\bf R_{1,2}} \equiv {\bf R_1}$ or $\bf R_2$, to the SM lepton doublet $L$ or the singlet $\nu_R$, one can attach $k$ Higgs fields as shown in Fig.~\ref{fig:mechanism}\,(b). Here $k=R_{1,2}-2$ to connect to $L$ and $k=R_{1,2}-1$ for $\nu_R$. It is important to realize that increasing the dimension of the representation $\bf R_{1,2}$ of the new fermion will also increase the EFT order of the effective operator (more SM Higgses), leading to suppressed magnetic moment.  

\section{Minimal UV Realization}
We now open up the effective vertex shown in Fig.~\ref{fig:mechanism} by constructing an explicit UV completion at the leading EFT order, i.e., dimension 6 (7) for the Dirac (Majorana) magnetic moment, at one-loop order. The resulting topology, shown in Fig.~\ref{fig:topology}, is obtained by setting $\bf R_1=3$ with a new heavy vector-like fermion $\Psi\sim\qn{1}{3}{0}$ and $\bf R_2=1$ with $\nu_R$ or $N\sim\qn{1}{1}{0}$, $n=0$ and $k=1$. The relevant Yukawa interaction of the fields $\{\Psi, L, H \}$ is 
\begin{equation}\label{eq:zero}
    y_{\Psi}\bar L \tau^a_{} \widetilde{H}  \Psi^a  + \text{h.c.}\,.
\end{equation}
Since $\Psi$ is heavy, the diagram in Fig.~\ref{fig:topology} is not one-light-particle irreducible (1LPI)\footnote{The presence of these bridge topologies were highlighted in the context of the muon anomalous magnetic moment \cite{Arkani-Hamed:2021xlp,DelleRose:2022ygn,Guedes:2022cfy,Crivellin:2023xbu}.}, and thus match directly to the dipole operator and/or the mass term after integrating out heavy degrees of freedom. 
To interact with the SM singlet $\nu_R$, the scalar $S$ and the fermion $F$ in the loop must have the same quantum numbers.
In this case, the Lagrangian terms for the two vertices in the loop (cf. Fig.~\ref{fig:topology}\,(a)) can be written as
\begin{equation} \label{eq:verticesinloop}
   \mathcal L \supset y_T\, \bar F_k  [T^{a}]^k_{l} S^{l} \Psi^a   
   +    y_\delta \,\delta^k_{l}\bar{F}_k S^{l} \nu_R     
   +  \text{h.c.}\,,
\end{equation}
where $SU(3)_c$ indices (if present) are implicit, and $T^a$ denote the $SU(2)_L$ generators in the representation of $S,F$. For instance, if the fields $(S,F)$ are adjoints of $SU(2)$, then $[T^a_A]^k_{l} = i \epsilon_{akl}$, where $\epsilon$ is the fully antisymmetric tensor. The chiral singlet field $\nu_R\sim\qn{1}{1}{0}$ in Eq.~\eqref{eq:verticesinloop} may be replaced by a heavy vector-like singlet $N\sim\qn{1}{1}{0})$ to construct the topology for the Majorana case, shown in Fig.~\ref{fig:topology}\,(b)\footnote{There is an additional topology with fourplet fermion in the Majorana case, see Appendix~\ref{app:leadingEFT} for details.}. 
Since $SU(2)_L$ generators are traceless (irrespective of the representation), it is clear that the product of the two vertices in Eq.~\eqref{eq:verticesinloop} appearing in the loop is zero, \textit{i.e.,} 
\begin{equation}\label{eq:zeroAMM}
    \delta^l_{k}[T^a]^k_{l} = \text{Tr}[T^a]=0.
\end{equation}
Consequently, the mass contribution from these topologies exactly vanishes. 
The contribution to the neutrino magnetic moment from the $U(1)_Y$ gauge boson, $B_\mu$, also vanishes for the same reason. On the other hand, the $W^3_\mu$ vertex comes with the third $SU(2)_L$ generator\footnote{The Lagrangian of the fields ($B_\mu$, $W_\mu$, $F$) read $g_1 \mathsf{y}_F \bar{F} \gamma^\mu F B_\mu + g_2 [T^a]^k_{l} \bar{F}_k \gamma^\mu F^{l} W_\mu^a $, where $\mathsf{y}_F$ is the hypercharge of F.} yielding a non-vanishing contribution proportional to
\begin{equation}\label{eq:nonzeroAMM}
   [T^a_{\bf R}]^k_{l} [T^3_{\bf R}]^{k'}_{k} \delta^l_{k'} = \text{Tr}[T^a_{\bf R} T^3_{\bf R}]=T({\bf R})\ \delta^{a3} \, ,
\end{equation}
 where $T({\bf R})$ is the Dynkin index in the $\bf R$ representation, \textit{i.e.,} $T({\bf R})=1/2, 2, 5...$ for the fundamental, the adjoint, the quadruplet representations and so on.
The details on the leading EFT order topologies including relevant loop functions and model-dependent coefficients are provided in the next section as well as in Appendix~\ref{app:Alltop}. The matching results were obtained using \texttt{Matchete} \cite{Fuentes-Martin:2022jrf} and the notebooks are provided as supplementary materials.
 
\subsection{UV--completion of Dirac \texorpdfstring{\pmb{$\mu_\nu^D$}}{Nu}}
\label{dirac_model} 
\setlength{\tabcolsep}{3pt}
\renewcommand{\arraystretch}{1.4}
\begin{table*}[]
    \centering
    \begin{tabular}{|c|c|c|c|c|c|c|c|}
    \hline
     Case & $SU(3)\times SU(2)\times U(1)$  & $U(1)_L$& Relevant Lagrangian ${\cal L}$  & $m_\nu $ [$\frac{\mu_\nu}{\mu_B m_e}$] & Enhanced $\mu_\nu ?$  \\[2pt] \hline \hline
     a & $\Psi \sim\qn{1}{3}{0}$ & $1$ & ${\cal L}_a = {\cal L}_{\rm SM} + \bar{\Psi}_R \Psi_L + \bar L \tilde H \Psi_R  + \bar L \tilde H \nu_R $ &  --  & \xmark \\[4pt]
     \hline 
     b & $F\sim\qn{1}{2}{1/2}$ & $1$ & ${\cal L}_a + \bar{F}_R F_L + \bar{F}_R H \Psi_L  + \bar{F}_L H \Psi_R  + \bar{F}_L H \nu_R $ &   $-\frac{3}{2} m_\Psi^2-4\pi^2v^2$ & \xmark \\[2pt]
     \hline
     c & $S\sim \qn{1}{3}{0}$ & 0 & ${\cal L}_a + \bar{\Psi}_R S \Psi_L   + \bar{\Psi}_L S \nu_R   + H^\dagger H S$ &  $-\frac{1}{4}|y_\Psi|^2 v^2$ & \cmark \\[2pt]
     \hline 
   d & $\{S,F\}\sim\qn{1}{4}{3/2}$ & \{0, 1\} & ${\cal L}_a + \bar{F}_R F_L + \bar{F}_R S \Psi_L + \bar{ F}_L S \Psi_R + \bar{F}_L S \nu_R   + S^\dagger H^3 $ & 0 & \cmark \\ \hline
   e & $\{S,F\}\sim({\bf 1},{\bf 2},1/2,-)$ & \{0, 1\} & ${\cal L}_a + \bar{F}_R F_L + \bar{F}_R S \Psi_L  + \bar{F}_L S \Psi_R+ \bar{F}_L S \nu_R      + H^\dagger S S^\dagger H$ & $ 0$& \cmark \\ \hline
    \end{tabular}
    \caption{Quantum numbers for minimal models that give rise to the topology of Fig.~\ref{fig:topology}(a) for Dirac neutrino magnetic moments. For the Majorana case, replacing $\nu_R$ with a vector-like singlet $N\sim\qn{1}{1}{0}$ yields the topology of Fig.~\ref{fig:topology}(b). Case (b)--(e) include the particle from (a), i.e., $\Psi\sim\qn{1}{3}{0}$.
}
    \label{tab:quantum_numbers}
\end{table*}

To realize a Dirac neutrino magnetic moment, we start by imposing a global lepton number symmetry, thereby forbidding Majorana mass term for $\nu_R$. Neutrinos acquire a Dirac mass of order \( 0.1 \, {\rm eV} \)  through the Yukawa interaction Eq.~\ref{eq:mass_operator_Dirac}, requiring $Y_{\rm Dirac} \sim10^{-12}$. In the minimal setup with only SM$+ \nu_R$ fields,  $\mu_\nu^D$ is proportional to the neutrino mass and is given by
$\mu_\nu^D = \frac{3e G_F}{8\sqrt{2} \pi^2} m_\nu \simeq 3 \times 10^{-19}\, \mu_B \left[ \frac{m_\nu}{\text{eV}} \right] \cite{Fujikawa:1980yx}$,
which lies seven orders of magnitude below the current experimental sensitivity. However, as discussed above, the topology in Fig.~\ref{fig:topology}\,(a) can induce a non-zero Dirac neutrino magnetic moment without inducing the corresponding mass term.

A broad class of such models can be constructed by introducing a new triplet fermion $\Psi\sim\qn{1}{3}{0}$, together with a scalar field \( S \sim \qn{a}{j}{q} \) and a fermion field \( F \sim \qn{a}{j}{q} \) with the same representations under the SM gauge group. $S,F$ can be either SM fields or new fields. The quantum numbers and relevant Lagrangian for minimal UV completions that generate the topology of Fig.~\ref{fig:topology}\,(a) are listed in Table~\ref{tab:quantum_numbers}. We begin with case (a), which contains only the field $\Psi$, which acts as the bridge between $L$ and the loop. In this case, $S$ is identified with the SM Higgs and $F$ with the SM lepton doublet, and the induced magnetic moment is proportional to the negligible Dirac Yukawa $Y_{\rm Dirac}$. The next minimal scenario involves introducing only one new field on top of $\Psi$. For example, in case (b) of Table~\ref{tab:quantum_numbers}, the scalar field $S$ is identified with the SM Higgs. In this scenario, although the loop contribution to the neutrino mass from the diagram in Fig.~\ref{fig:topology} vanishes, both tree-level and loop-level diagrams (with different topologies) at dimension-6 contribute to the mass, involving the same set of Yukawa couplings $y_{T}^* y_\Psi y_\delta$. As a result, it is not possible to enhance the magnetic moment without simultaneously generating a large neutrino mass.  

The first viable minimal model is presented in case (c), where two new fields are added: a vector-like fermion triplet \( F \equiv \Psi \sim \qn{1}{3}{0} \), and a real scalar triplet \( S \sim \qn{1}{3}{0} \). The most relevant Yukawa interactions are given by Eq.~\eqref{eq:zero} and  Eq.~\eqref{eq:verticesinloop}, and there is a mass term for $\Psi$. The scalar potential involving the fields ($H$,$S$) is
\begin{align}
    V &= - \mu_H^2 (H^\dagger H) + \mu_S^2  (S^a S^a) + \lambda_H (H^\dagger H)^2 + \lambda_{S} (S^a S^a)^2 \notag \\
    &\quad + \lambda_{HS} (H^\dagger H) (S^a S^a) + \mu \, H^\dagger \tau^a H S^a \, .
    \label{eq:tripletpot}
\end{align}
In the limit $\mu \to 0$, the scalar potential has an enhanced global symmetry $O(4)_H \times O(3)_S$ symmetry. Therefore, it is technically natural to take $\mu \ll 1$.   

After integrating out the heavy fields $(S,\Psi)$, Fig.~\ref{fig:topology}\,(a) generates only the dimension-6 operator $O_{\nu W}$ of Eq.~\eqref{eq:moment_operator_Dirac} with Wilson coefficient
\begin{equation} \label{eq:dipolecontribution}
    C_{\nu W} = \frac{y_\Psi y_\delta y_T^*}{
    16 \pi^2 } \frac{m_S^2 \left( 1- \log\left(m_S^2/m_\Psi^2\right) \right) -m_\Psi^2}{(m_\Psi^2-m_S^2)^2}\, .
\end{equation}
After electroweak symmetry breaking, the neutrino magnetic moment $\mu_\nu^D$ is 
\begin{equation}
    \frac{\mu_\nu^D}{\mu_B} = 2\sqrt 2\ m_e v\ C_{\nu W} \, .
    \label{eq:mass_dim6}
\end{equation}
In addition to Fig.~\ref{fig:topology}\,(a), there are other dimension-6 topologies contributing to $\mu_\nu^D$ through $C_{\nu B}$ and/or $C_{\nu W}$. However, all such diagrams are either proportional to the Dirac Yukawa coupling $Y_{\rm Dirac}$ or the cubic coupling $\mu$. Hence, in the limit $Y_{\rm Dirac}, \mu \to 0$, their contributions are subleading compared to the dominant contribution from Fig.~\ref{fig:topology}\,(a). 

We now discuss the generation of Dirac neutrino mass arising from dimension-6 operator of Eq.~\eqref{eq:mass_operator_Dirac}, where $C_{\nu H}$ after integrating out the heavy fields receives a one-loop contribution $C_{\nu H} = 2 C_{\nu W} |y_\Psi|^2$. 
The resulting neutrino mass is   
\begin{equation}
    m_\nu^{\rm loop} = -\frac{\mu_\nu^D}{\mu_B} \frac{|y_\Psi|^2 v^2}{4 m_e} \, . 
    \label{eq:dim6_loop_mass}
\end{equation}
Here the mass is suppressed by $\left(|y_\Psi|v/\Lambda\right)^2$, where $\Lambda\sim m_\Psi,m_S$, with respect to the relation given in Eq.~\eqref{eq:numass_NMM}. If one considers a non-minimal extension with three new particles as shown in case (d) and (e) of Table~\ref{tab:quantum_numbers}, the neutrino masses remain zero at dimension 6 both at tree level and loop level. Case (d) demonstrates possible higher representations under the SM gauge group, whereas case (e) shows a model with an extra discrete symmetry under which the loop particles are odd while all the SM fermions are even. In this set-up, the lightest particle can be a dark matter candidate \cite{Cirelli:2005uq,LopezHonorez:2006gr,Restrepo:2013aga}. 
  
Assuming new physics scale $m_\Psi, m_S \sim 1$ TeV, Yukawa couplings $y_\delta, y_T \sim 1$ and $y_\Psi \sim 10^{-3}$, the induced neutrino magnetic moment in case (c) can reach $\mu_\nu^D \sim  10^{-12}\ \mu_B$, within the sensitivity of future experiments. In non-minimal scenarios such as cases (d) and (e), the magnetic moment can be much larger, as the neutrino mass remain zero at least up to one-loop generated dimension-6 operators.

However, this setup does not a priori escape the model-independent bound, $\mu_\nu^D \lesssim 10^{-15} \mu_B$, arising from the mixing of the dimension-6 operators $O_{\nu W}$ into $O_{\nu H}$ via renormalization group running from the high scale $\Lambda \sim 1$ TeV to the electroweak scale~\cite{Bell:2005kz}.
Current experimental bounds thus place meaningful constraints on these models.

\subsection{UV-completion of Majorana \texorpdfstring{\pmb{$\mu_\nu^M$}}{NMM}}
In this section, we show that the same mechanism can be extended to generate a Majorana neutrino transition magnetic moment, as illustrated in Fig.~\ref{fig:topology}\,(b). Here the right handed neutrino $\nu_R$ is replaced by a vector-like singlet fermion $N\sim\qn{1}{1}{0}$. Due to the underlying mechanism prescribed by Eq.~\eqref{eq:zeroAMM}, the contribution to the neutrino mass from Fig.~\ref{fig:topology}\,(b) vanishes, i.e.,  $C_{LH}^5 = 0$ in Eq.~\ref{eq:mass_operator_Majorana}, while a non-zero magnetic moment is generated. On the other hand, the dimension-7 mass operator in Eq.~\ref{eq:mass_operator_Majorana} may or may not be induced, depending on the quantum numbers of the new scalar and fermion fields $S$ and $F$. 
We present a concrete realization of this model by extending case (c) of Table.~\ref{tab:quantum_numbers} with a second Higgs doublet $\Phi\sim\qn{1}{2}{1/2}$ carrying lepton number $L =2$. The most general Yukawa interactions and mass terms of this model are
\begin{align}
    {\cal L} &=~ 
    - m_N\bar{N}_R N_L 
    - m_\Psi \bar{\Psi}_R \Psi_L 
    + y_{N} \bar L \tilde H {N}_R  
     + y_\Psi \bar{L} \tau^a \tilde H \Psi_R^a   \notag \\ 
    & + y_N' \overline{L^c} \tilde\Phi^\dagger N_L 
     + y_\Psi' \overline{L^c} \tau^a \tilde\Phi^\dagger \Psi_L^a 
      - y_T \,if^{abc}\bar{\Psi}_R^a S^b \Psi_L^c 
    \notag \\ 
    &+ y_{\delta} \bar{N}_R  S^a  \Psi_L^a  
    + y_{\delta}' \bar{N}_L  S^a \Psi_R^a
    + \text{h.c.}  \, 
\end{align}
The relevant and non-trivial terms in the scalar potential are
\begin{equation}
    V \supset \lambda_{SH1} H^\dagger \tau^a H S^a + \mu^2  \Phi^\dagger  H+ \text{h.c.} \, ,
\end{equation}
with $\mu$ softly breaking lepton number by two units and enabling the generation of the desired magnetic moment diagram. Depending on the UV completion, these interactions may also induce a Majorana neutrino mass at dimension-7 or higher. It is worth noting that the soft-breaking term $\mu^2 H \Phi^\dagger$ can arise dynamically by promoting global lepton number $U(1)_L$ to a local gauge symmetry. 
After integrating out the heavy new fields and electroweak symmetry breaking, the magnetic moment for Majorana neutrino $\mu_\nu^M$ at one loop level is given by
\begin{equation}
    \frac{\mu_\nu^M}{\mu_B} = 4\, m_e v^2 C_{LHW} \, ,
\end{equation}
where 
\begin{equation}
    C_{LHW} =  \frac{ y_\Psi^* y_\delta y_T y_N' \theta_{\Phi H}}{ 16 \pi^2 }  \frac{m_\Psi^2 + m_S^2 \left(\log\left(m_S^2/m_\Psi^2\right) -1  \right) }{m_N (m_\Psi^2-m_S^2)^2}\, .
\end{equation}
Here $\theta_{\Phi H}$ is the mixing term between the SM Higgs field and $\Phi$ field and is of order $\mu^2/m_\Phi^2$. 
Similarly to the Dirac case, there is also a radiative correction to the neutrino mass given by
\begin{equation}
    m_\nu^{\rm loop} = - \frac{\mu_\nu^M}{\mu_B} \frac{|y_\Psi|^2 v^2}{ 8m_e} \, . 
    \label{eq:dim6_loop_mass_1}
\end{equation}
For Yukawa couplings $y_\delta, y_T, y_N \sim 1$ and $y_\Psi \sim 10^{-3}$, particle masses $m_\Psi, m_S \sim 1~\text{TeV}, m_N \sim 100~\text{GeV}$, and scalar mixing $\theta_{\Phi H} \sim 0.1$, the magnetic moment of a Majorana neutrino can reach $\mu_\nu^M \sim 10^{-12}\,\mu_B$. In non-minimal scenarios, where neutrino masses vanish exactly at dimensions 5 and 7, larger values of the $\mathcal{O}(10^{-9})\,\mu_B$ are achievable.
Unlike the Dirac case, model-independent RGE bounds on Majorana neutrino magnetic moments are weaker, typically $\mu_\nu^M < 10^{-10}\,\mu_B$~\cite{Davidson:2005cs,Bell:2006wi}. 


\section{Conclusion}
We have proposed a new mechanism for generating large neutrino magnetic moments for both Dirac and Majorana neutrinos, while naturally suppressing the associated mass contributions without fine-tuning. 
The key lies in the careful choice of $SU(2)$ representations in a bridge topology \textit{i.e.,} where the neutrino connects to the loop via one or more heavy fermion(s). At leading order in the EFT, the heavy fermion is chosen in the adjoint representation of $SU(2)_L$. In all the cases discussed, the magnetic moment arises at one loop, while the corresponding mass diagram --- obtained by removing the photon --- vanishes due to $SU(2)$ invariance.

While our analysis focused on a minimal realization within the SM gauge group, the mechanism is broadly applicable. This opens new directions for constructing models with observable neutrino magnetic moments, both within and beyond the SM gauge group, potentially testable in current and future experiments. Moreover, by suitably assigning the quantum numbers of the particles inside the loop, the same model could also address other open questions, such as flavor anomalies, dark matter, and baryogenesis, offering a unified connection to neutrino properties. This highlights the potential of neutrino magnetic moments as a powerful probe of new physics.

\begin{acknowledgments}
We thank K S Babu, Aneesh Manohar and R N Mohapatra for useful discussions and comments on the manuscript. We gratefully acknowledge the hospitality of CERN during the final stages of this work. JP is supported by the U.S. Department of Energy (DOE) under award numbers DE-SC0009919. 

\end{acknowledgments}

\newpage
\appendix
\renewcommand{\thesection}{\Roman{section}}
 \renewcommand{\thesubsection}{\roman{subsection}}
  \renewcommand{\thesubsubsection}{\roman{subsubsection}}
\addcontentsline{toc}{section}{Appendices}

\section{Leading EFT order topologies}\label{app:leadingEFT}
In this section we provide all the topologies for both Dirac and Majorana magnetic moment at the leading EFT order, i.e., we focus on dimension-6 operators for Dirac and dimension-7 operators for Majorana neutrinos. 
\label{app:Alltop}

\subsubsection{Dirac topologies}\label{app:Diractop}
\begin{figure}[h]
    \begin{tikzpicture}
    \tikzset{
                fermion/.style={draw=black, line width=1.0pt, postaction={decorate}, 
                },
                 fermion_no_arrow/.style={draw=red, line width=1.0pt},
                scalar/.style={draw=black, dashed, line width=1.0pt},
                photon/.style={decorate, decoration={snake, amplitude=2pt, segment length=6pt}, draw=black, line width=1.0pt},
                scalar/.style={draw=black, dashed, line width=1.0pt}
            }
    \draw[fermion] (-2,0) node[left, below] {$L$} -- (-1.3,0) node[midway, above] {\color{gray}2}  ;
    \draw[fermion_no_arrow] (-1.3,0) -- (-0.5,0) node[midway, above] {\color{gray}1,3}  ;
    \draw[fill=gray!30] (0,0) circle (0.5);
    \draw[fermion] (0.5,0)  -- (1.3,0) node[midway, above] {\color{gray}1} node[right, below] {$\nu_R$};
    \draw[photon] (0.45,0.45) -- (1,1) node[midway, above left] {\color{gray}3} node[above] {$W$};
    \draw[scalar] (-1.3,0) -- (-1.3,1) node[midway, above left] {\color{gray}2} node[above] {$H$};
    \node at (0,-0.9) {(a)};
    \end{tikzpicture}
    ~~~~~~~~~~~~~~
    \\
    \begin{tikzpicture}
   \tikzset{
                fermion/.style={draw=black, line width=1.0pt, postaction={decorate}, 
                },
                photon/.style={decorate, decoration={snake, amplitude=2pt, segment length=6pt}, draw=black, line width=1.0pt},
                scalar/.style={draw=black, dashed, line width=1.0pt}
            }
    \draw[fermion] (-1.3,0) node[left, below] {$L$} -- (-0.5,0) node[midway, above] {\color{gray}2}  ;
    \draw[fill=gray!30] (0,0) circle (0.5);
    \draw[fermion] (0.5,0)  -- (1.3,0) node[midway, above] {\color{gray}1} node[right, below] {$\nu_R$};
    \draw[photon] (0.45,0.45) -- (1,1) node[midway, above left] {\color{gray}3} node[above] {$W$};
    \draw[scalar] (-0.35,0.35) -- (-1,1) node[midway, above right] {\color{gray}2} node[above] {$H$};
    \node at (0,-0.9) {(b)};
    \end{tikzpicture}
    \hfill
    \begin{tikzpicture}
    \tikzset{
                fermion/.style={draw=black, line width=1.0pt, postaction={decorate}, 
                },
                 fermion_no_arrow/.style={draw=red, line width=1.0pt},
                scalar/.style={draw=black, dashed, line width=1.0pt},
                photon/.style={decorate, decoration={snake, amplitude=2pt, segment length=6pt}, draw=black, line width=1.0pt},
                scalar/.style={draw=black, dashed, line width=1.0pt}
            }
    \draw[fermion] (-1.3,0) node[left, below] {$L$} -- (-0.5,0) node[midway, above] {\color{gray}2}  ;
    \draw[fermion_no_arrow] (0.5,0) -- (1.3,0) node[midway, above] {\color{gray}2}  ;
    \draw[fill=gray!30] (0,0) circle (0.5);
    \draw[fermion] (1.3,0)  -- (2,0) node[midway, above] {\color{gray}1} node[right, below] {$\nu_R$};
    \draw[photon] (-0.45,0.45) -- (-1,1) node[midway, above right] {\color{gray}3} node[above] {$W$};
    \draw[scalar] (1.3,0) -- (1.3,1) node[midway, above left] {\color{gray}2} node[above] {$H$};
    \node at (0,-0.9) {(c)};
    \end{tikzpicture}
\caption{All topologies potentially generating the Dirac mass at dimension 4 and the magnetic moment at dimension 6 (by attaching the $W_\mu$ leg) of the neutrinos. $SU(2)_L$ representations are shown in gray.    \label{fig:othertopDirac}}
\end{figure}
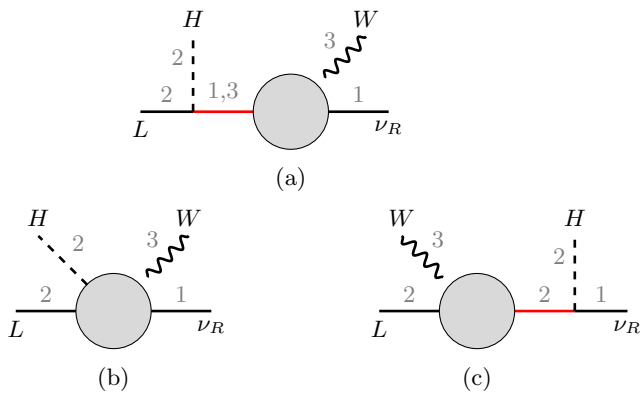

Fig.~\ref{fig:othertopDirac} shows all the relevant topologies contributing to the Dirac neutrino mass at dimension-4 and/or to the magnetic moment at dimension-6. Among them, Fig.~\ref{fig:othertopDirac}\,(a) and (c) represent bridge topologies, where a heavy vector-like fermion connects the loop to an external neutrino line via a Higgs insertion. 
\setlength{\tabcolsep}{0pt}
\renewcommand{\arraystretch}{1.2}
\begin{table}[h]
    \centering
    \begin{tabular}{|c|c|c|c|}
    \hline
     ~Topology~ & ~Rep. in loop~  & ~$\supset 1$~ & ~$\supset 3$ ~ \\ \hline
     \multirow{2}{*}{(a)} & $\bf 1 \otimes 1$ & \cmark & \xmark 
     \\ \cline{2-4}
      & \cellcolor{green!25}$\bf 3 \otimes 1$ & \cellcolor{green!25}\xmark & \cellcolor{green!25}\cmark \\
      
     \hline 
     (b) & $\bf 2 \otimes 2 \otimes 1$ & \cmark & \cmark 
     \\ \hline 
     (c) & $\bf 2 \otimes 2 $ & \cmark & \cmark 
     \\ \hline 
    \end{tabular}
    \caption{$SU(2)_L$ tensor products of the representations in the topologies of Fig.~\ref{fig:othertopDirac}. Highlighted in green is the topology with the potential to generate the dipole without the mass. }
    \label{tab:Diractop}
\end{table}

As can be seen from Table.~\ref{tab:Diractop}, topology~\ref{fig:othertopDirac}\,(a) with $SU(2)_L$ triplet fermion only exists when the $W_\mu$ line is connected, i.e., $\bf 3 \otimes 3 \supset 1$ allows the dipole diagram (with $W_\mu$), but not the mass diagram (without it). For topologies (b)  and (c) of Fig.~\ref{fig:othertopDirac}, invariant contractions exist regardless of whether $W_\mu$ leg is attached, \textit{i.e.,} $\bf 2 \otimes 2=3\oplus1$ allowing for both mass and dipole operators. 

\subsubsection{Majorana topologies}\label{app:Majoranatop}
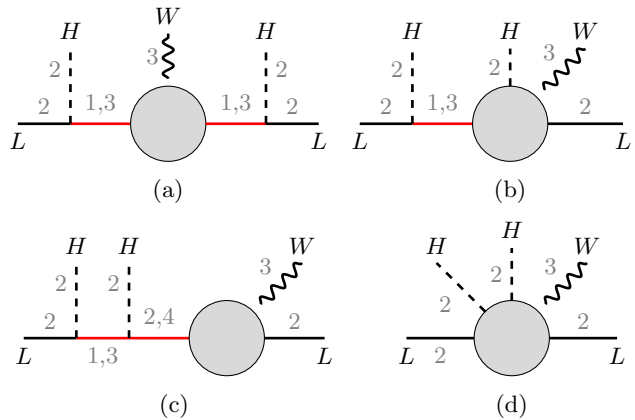
\begin{figure}[h]
    \begin{tikzpicture}
    \tikzset{
                fermion/.style={draw=black, line width=1.0pt, postaction={decorate}, 
                },
                 fermion_no_arrow/.style={draw=red, line width=1.0pt},
                scalar/.style={draw=black, dashed, line width=1.0pt},
                photon/.style={decorate, decoration={snake, amplitude=2pt, segment length=6pt}, draw=black, line width=1.0pt},
                scalar/.style={draw=black, dashed, line width=1.0pt}
            }
    \draw[fermion] (-2,0) node[left, below] {$L$} -- (-1.3,0) node[midway, above] {\color{gray}2}  ;
    \draw[fermion_no_arrow] (-1.3,0) -- (-0.5,0) node[midway, above] {\color{gray}1,3}  ;
    \draw[fill=gray!30] (0,0) circle (0.5);
    \draw[fermion_no_arrow] (0.5,0) -- (1.3,0) node[midway, above] {\color{gray}1,3}  ;
    \draw[fermion] (1.3,0)  -- (2,0) node[midway, above] {\color{gray}2} node[right, below] {$L$};
    \draw[photon] (0,0.6) -- (0,1.2) node[midway, left] {\color{gray}3} node[above] {$W$};
    \draw[scalar] (-1.3,0) -- (-1.3,1) node[midway, above left] {\color{gray}2} node[above] {$H$};
    \draw[scalar] (1.3,0) -- (1.3,1) node[midway, above right] {\color{gray}2} node[above] {$H$};
    \node at (0,-0.9) {(a)};
    \end{tikzpicture}
    \begin{tikzpicture}
    \tikzset{
                fermion/.style={draw=black, line width=1.0pt, postaction={decorate}, 
                },
                 fermion_no_arrow/.style={draw=red, line width=1.0pt},
                scalar/.style={draw=black, dashed, line width=1.0pt},
                photon/.style={decorate, decoration={snake, amplitude=2pt, segment length=6pt}, draw=black, line width=1.0pt},
                scalar/.style={draw=black, dashed, line width=1.0pt}
            }
    \draw[fermion] (-2,0) node[left, below] {$L$} -- (-1.3,0) node[midway, above] {\color{gray}2}  ;
    \draw[fermion_no_arrow] (-1.3,0) -- (-0.5,0) node[midway, above] {\color{gray}1,3}  ;
    \draw[fill=gray!30] (0,0) circle (0.5);
    \draw[fermion] (.5,0)  -- (1.5,0) node[midway, above] {\color{gray}2} node[right, below] {$L$};
    \draw[photon] (0.45,0.45) -- (1,1) node[midway, above left] {\color{gray}3} node[above] {$W$};
    \draw[scalar] (-1.3,0) -- (-1.3,1) node[midway, above left] {\color{gray}2} node[above] {$H$};
    \draw[scalar] (0,0.5) -- (0,1) node[midway,  left] {\color{gray}2} node[above] {$H$};
    \node at (0,-0.9) {(b)};
    \end{tikzpicture}
      \begin{tikzpicture}
    \tikzset{
                fermion/.style={draw=black, line width=1.0pt, postaction={decorate}, 
                },
                 fermion_no_arrow/.style={draw=red, line width=1.0pt},
                scalar/.style={draw=black, dashed, line width=1.0pt},
                photon/.style={decorate, decoration={snake, amplitude=2pt, segment length=6pt}, draw=black, line width=1.0pt},
                scalar/.style={draw=black, dashed, line width=1.0pt}
            }
    \draw[fermion] (-2.7,0) node[left, below] {$L$} -- (-2,0) node[midway, above] {\color{gray}2}  ;
    \draw[fermion_no_arrow] (-2,0) -- (-1.3,0) node[midway, below] 
    {\color{gray}1,3}  ;
    \draw[fermion_no_arrow] (-1.3,0) -- (-0.5,0) node[midway, above] {\color{gray}2,4}  ;
    \draw[fill=gray!30] (0,0) circle (0.5);
    \draw[fermion] (.5,0)  -- (1.3,0) node[midway, above] {\color{gray}2} node[right, below] {$L$};
    \draw[photon] (0.45,0.45) -- (1,1) node[midway, above left] {\color{gray}3} node[above] {$W$};
    \draw[scalar] (-1.3,0) -- (-1.3,1) node[midway, above left] {\color{gray}2} node[above] {$H$};
    \draw[scalar] (-2,0) -- (-2,1) node[midway,  above left] {\color{gray}2} node[above] {$H$};
    \node at (-.7,-.9) {(c)};
    \end{tikzpicture}
    ~~~~
       \begin{tikzpicture}
    \tikzset{
                fermion/.style={draw=black, line width=1.0pt, postaction={decorate}, 
                },
                 fermion_no_arrow/.style={draw=red, line width=1.0pt},
                scalar/.style={draw=black, dashed, line width=1.0pt},
                photon/.style={decorate, decoration={snake, amplitude=2pt, segment length=6pt}, draw=black, line width=1.0pt},
                scalar/.style={draw=black, dashed, line width=1.0pt}
            }
    \draw[fermion] (-1.4,0) node[left, below] {$L$} -- (-.5,0) node[midway, below] {\color{gray}2}  ;
    \draw[fill=gray!30] (0,0) circle (0.5);
    \draw[fermion] (.5,0)  -- (1.4,0) node[midway, above] {\color{gray}2} node[right, below] {$L$};
    \draw[photon] (0.45,0.45) -- (1,1) node[midway, above left] {\color{gray}3} node[above] {$W$};
    \draw[scalar] (-0.35,0.35) -- (-1,1) node[midway, below left] {\color{gray}2} node[above] {$H$};
    \draw[scalar] (0,0.5) -- (0,1.2) node[midway,  left] {\color{gray}2} node[above] {$H$};
    \node at (0,-0.9) {(d)};
    \end{tikzpicture}
\caption{All topologies potentially generating the Majorana mass at dimension 5 and the magnetic moment at dimension 7 (by attaching the $W_\mu$ leg) of the neutrinos. $SU(2)_L$ representations are shown in gray.   \label{fig:othertopMajorana}}
\end{figure}

Fig.~\ref{fig:othertopMajorana} displays all the topologies that contribute to the Majorana neutrino mass at dimension-5 and/or to the magnetic moment at dimension-7. Fig.~\ref{fig:othertopMajorana}\,(a), (b) and (c) corresponds to bridge topologies.
Contrary to the Dirac case, there is no distinction between the incoming and outgoing fermion leg since it is the same particle $\ell$. As a consequence each graph can be mirrored. 

\setlength{\tabcolsep}{0pt}
\renewcommand{\arraystretch}{1.2}
\begin{table}[h]
    \centering
    \begin{tabular}{|c|c|c|c|}
    \hline
     ~Topology~ & ~Rep. in loop~  & ~$\supset 1$~ & ~$\supset 3$~  \\ \hline
     \multirow{3}{*}{(a)} & $\bf 1 \otimes 1$ & \cmark & \xmark 
     \\ \cline{2-4}
     & \cellcolor{green!25}$\bf 1 \otimes 3$ or $\bf 3\otimes 1$ &\cellcolor{green!25} \xmark & \cellcolor{green!25}\cmark 
     \\ \cline{2-4}
      & $\bf 3 \otimes 3$ & \cmark & \cmark 
      \\\hline 
     \multirow{2}{*}{(b)} & $\bf 1 \otimes 2 \otimes 2 $ & \cmark & \cmark 
     \\ \cline{2-4}
     & $\bf 3 \otimes 2 \otimes 2 $ & \cmark & \cmark 
     \\ \hline
     \multirow{2}{*}{(c)} & $\bf  2 \otimes 2 $ & \cmark & \cmark 
     \\ \cline{2-4}
     & \cellcolor{green!25}$\bf 4 \otimes 2 $ & \cellcolor{green!25}\xmark & \cellcolor{green!25}\cmark 
     \\ \hline
     (d) & $\bf 2 \otimes 2\otimes 2\otimes 2 $ & \cmark & \cmark 
     \\ \hline 
    \end{tabular}
    \caption{$SU(2)_L$ tensor products of the representations in the topologies of Fig.~\ref{fig:othertopMajorana}. Highlighted in green are the topologies with the potential to generate the dipole without the mass. }
    \label{tab:Majoranatop}
\end{table}
As summarized in Table.~\ref{tab:Majoranatop}, Fig.~\ref{fig:othertopMajorana}\,(a) with one triplet vector-like lepton and Fig.~\ref{fig:othertopMajorana} (c) with a triplet and a fourplet vector-like leptons only exist when the $W_\mu$ line is connected. This implies that the corresponding Majorana mass diagram --- obtained by removing the photon --- does not exist, and the magnetic moment is generated without inducing a mass.

\section{Example models for Dirac case}
\label{app:models}
For completeness, we now provide further details on models (b) and (c) of Table~\ref{tab:quantum_numbers}. 

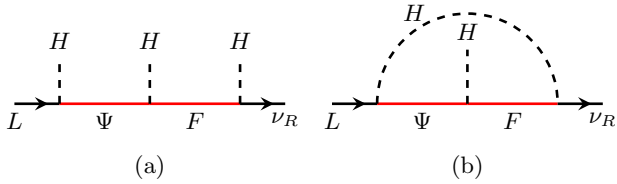
\begin{figure}[!h]
    \begin{tikzpicture}[scale=1.2]
            \tikzset{
                fermion/.style={draw=black, line width=1.0pt, postaction={decorate}, decoration={markings,mark=at position .75 with {\arrow[scale=1.1, line width=1.0pt]{stealth}}}},
                fermion_no_arrow/.style={draw=red, line width=1.0pt},
                scalar/.style={draw=black, dashed, line width=1.0pt},
                scalar_red/.style={draw=red, dashed, line width=1.0pt},
                photon/.style={decorate, decoration={snake, amplitude=2pt, segment length=6pt}, draw=black, line width=1.0pt}
            }
            \draw[fermion] (-2.5,0) node[left, below]{$L$}  -- (-2,0)  ;
            \draw[fermion_no_arrow] (-2,0) -- (-1,0) node[midway, below]{\textcolor{black}{$\Psi$}} ;
            \draw[fermion_no_arrow] (-1,0) -- (0,0) node[midway, below] {\textcolor{black}{$F$}};
            \draw[fermion] (0.,0) -- (0.5,0) node[right, below] {$\nu_R$};
            
            \draw[scalar] (-2,0) -- (-2,0.5) node[above] {$H$};
            \draw[scalar] (-1,0) -- (-1,0.5) node[above] {$H$};
            \draw[scalar] (0,0) -- (0,0.5) node[above] {$H$};
            \node at (-1,-0.7) {(a)};
    \end{tikzpicture}
    \begin{tikzpicture}[scale=1.2]
            \tikzset{
                fermion/.style={draw=black, line width=1.0pt, postaction={decorate}, decoration={markings,mark=at position .75 with {\arrow[scale=1.1, line width=1.0pt]{stealth}}}},
                fermion_no_arrow/.style={draw=red, line width=1.0pt},
                scalar/.style={draw=black, dashed, line width=1.0pt},
                scalar_red/.style={draw=red, dashed, line width=1.0pt},
                photon/.style={decorate, decoration={snake, amplitude=2pt, segment length=6pt}, draw=black, line width=1.0pt}
            }
            \draw[fermion] (-2.5,0) node[left, below]{$L$}  -- (-2,0)  ;
            \draw[fermion_no_arrow] (-2,0) -- (-1,0) node[midway, below]{\textcolor{black}{$\Psi$}} ;
            \draw[fermion_no_arrow] (-1,0) -- (0,0) node[midway, below] {\textcolor{black}{$F$}};
            \draw[fermion] (0,0) -- (.5,0) node[right, below] {$\nu_R$};
            
            \draw[scalar] (-2,0) arc[start angle=180, end angle=0, radius=1] node[pos=0.3, above] {$H$};
            \draw[scalar] (-1,0) -- (-1,0.6) node[above] {$H$};
            \node at (-1,-0.7) {(b)};
    \end{tikzpicture}
\caption{Tree-level (loop-level) dimension-6 (dimension-4) graph for model (b) of Table.~\ref{tab:quantum_numbers}. \label{fig:doubletlooptopology}}
\end{figure}

\subsubsection{Case (b) with \texorpdfstring{$F\sim \qn{1}{2}{1/2}$ and $S\equiv H \sim \qn{1}{2}{1/2}$}{ca}}
This is the minimal realization in which the scalar $S$ running in the loop is identified with the SM Higgs doublet. The Lagrangian is augmented by the following interactions
\begin{align}
    {\cal L} &\supset  y_{T}  \bar{F}_R \tau^a H \Psi_L^a + y_{T}'  \bar F_L \tau^a H   \Psi_R^a  \nonumber \\ 
    & + y_\delta \bar{F}_{L} H \nu_R 
    + {\rm h.c.} 
\end{align}
additionally to Eq.~\eqref{eq:zero}, along with the kinetic and mass terms for the vector-like fermions $\Psi$ and $F$.

In this case, although the contribution to the neutrino mass from the loop diagram in Fig.~\ref{fig:topology}\,(a) vanishes due to the proposed mechanism, additional contributions arises: tree-level mass term at dimension-6 from the diagram in Fig.~\ref{fig:doubletlooptopology}\,(a) and loop-level mass contribution at dimension-4 from a different topology involving the same particle content, shown in Fig.~\ref{fig:doubletlooptopology}\,(b). Both of these graphs involve the same combination of couplings, $y_{T}^* y_\Psi y_\delta$, implying that the magnetic moment and mass contribution are directly linked. As a result, it is not possible to generate a sizable neutrino magnetic moment without also inducing a large neutrino mass.

\subsubsection{Case (c) with \texorpdfstring{$F\equiv\Psi\sim \qn{1}{3}{0}$ and $S \sim \qn{1}{3}{0}$}{caseb}}
This case is a minimal working model (case (c) of Table~\ref{tab:quantum_numbers}), as discussed in the main text, that can enhance neutrino magnetic moments. Here we provide additional details and supporting diagrams associated with it. It only introduces an extra scalar on top of the fermion $\Psi$, which is also the same fermion running inside the loop. The relevant Yukawa interaction and the potential are given by Eq.~\eqref{eq:zero}, Eq.~\eqref{eq:verticesinloop}, and Eq.~\eqref{eq:tripletpot}. 
\begin{figure}[!t]
    \begin{tikzpicture}[scale=1.2]
            \tikzset{
                fermion/.style={draw=black, line width=1.0pt, postaction={decorate}, decoration={markings,mark=at position .75 with {\arrow[scale=1.1, line width=1.0pt]{stealth}}}},
                fermion_no_arrow/.style={draw=red, line width=1.0pt},
                scalar/.style={draw=black, dashed, line width=1.0pt},
                scalar_red/.style={draw=red, dashed, line width=1.0pt},
                photon/.style={decorate, decoration={snake, amplitude=2pt, segment length=6pt}, draw=black, line width=1.0pt}
            }
            \draw[fermion] (-2.5,0) node[left, below]{$L$}  -- (-1.75,0)  ;
            \draw[fermion_no_arrow] (-1.75,0) -- (-1,0) node[ below]{\textcolor{black}{$\Psi$}} ;
            \draw[fermion_no_arrow] (-1,0) -- (-.25,0)  ;
            \draw[fermion] (-.25,0) -- (.5,0) node[right, below] {$\nu_R$};
            
            \draw[scalar_red] (-1,0.75) arc[start angle=90, end angle=0, radius=.75] node[pos=0.6, above=.05in] {$S$};
            \draw[scalar] (-1.75,0) arc[start angle=180, end angle=90, radius=.75] node[pos=0.4, above=.05in] {$H$};
            \draw[scalar] (-1,.75) -- (-1,1.2) node[above] {$H$};
    \end{tikzpicture}
    \begin{tikzpicture}[scale=1.2]
            \tikzset{
                fermion/.style={draw=black, line width=1.0pt, postaction={decorate}, decoration={markings,mark=at position .75 with {\arrow[scale=1.1, line width=1.0pt]{stealth}}}},
                fermion_no_arrow/.style={draw=red, line width=1.0pt},
                scalar/.style={draw=black, dashed, line width=1.0pt},
                scalar_red/.style={draw=red, dashed, line width=1.0pt},
                photon/.style={decorate, decoration={snake, amplitude=2pt, segment length=6pt}, draw=black, line width=1.0pt}
            }
            \draw[fermion] (-2.5,0) node[left, below]{$L$}  -- (-1.75,0)  ;
            \draw[fermion_no_arrow] (-1.75,0) -- (-1,0) node[ below]{\textcolor{black}{$\Psi$}} ;
            \draw[fermion_no_arrow] (-1,0) -- (-.25,0)  ;
            \draw[fermion] (-.25,0) -- (.5,0) node[right, below] {$\nu_R$};
            
            \draw[scalar_red] (-0.25,0) -- (-0.25,.5) node[midway, left] {$S$};
            \draw[scalar] (-0.25,.5) -- (-.75,1) node[above left] {$H$};
            \draw[scalar] (-0.25,.5) -- (0.25,1) node[above right] {$H$};
            \draw[scalar] (-1.75,0) -- (-1.75,1) node[above] {$H$};
    \end{tikzpicture}
\caption{Feynman diagram that contributes to the mass at dimension-4 (dimension-6) at the loop- (tree-) level. \label{fig:dim4tripletlooptopology}}
\end{figure}
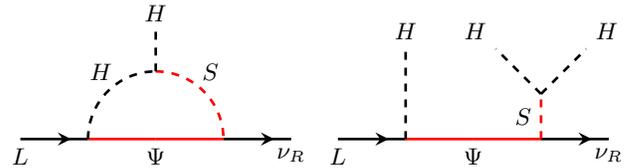
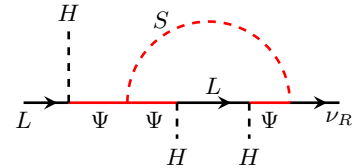
\begin{figure}[!t]
        \begin{tikzpicture}[scale=1.2]
            \tikzset{
                fermion/.style={draw=black, line width=1.0pt, postaction={decorate}, decoration={markings,mark=at position .75 with {\arrow[scale=1.1, line width=1.0pt]{stealth}}}},
                fermion_no_arrow/.style={draw=red, line width=1.0pt},
                scalar/.style={draw=black, dashed, line width=1.0pt},
                scalar_red/.style={draw=red, dashed, line width=1.0pt},
                photon/.style={decorate, decoration={snake, amplitude=2pt, segment length=6pt}, draw=black, line width=1.0pt}
            }
            \draw[fermion] (-2.5,0) node[left, below]{$L$}  -- (-2,0)  ;
            \draw[fermion_no_arrow] (-2,0) -- (-1.3,0) node[midway, below]{\textcolor{black}{$\Psi$}} ;
            \draw[fermion_no_arrow] (-1.3,0) -- (-.8,0) node[midway, below]{\textcolor{black}{$\Psi$}} ;
            \node at (-.4,0) [above] {$L$};
            \draw[fermion] (-.8,0)  -- (0,0)  ;
            \draw[fermion_no_arrow] (0,0) -- (0.45,0) node[midway, below]{\textcolor{black}{$\Psi$}} ;
            \draw[fermion] (0.45,0) -- (1.0,0) node[right, below] {$\nu_R$};
            
            \draw[scalar_red] (-1.35,0) arc[start angle=180, end angle=0, radius=0.9] node[pos=0.3, above] {$S$};
            \draw[scalar] (-2,0) -- (-2,0.8) node[above] {$H$};
            \draw[scalar] (0,0) -- (0,-0.4) node[below] {$H$};
            \draw[scalar] (-.8,0) -- (-.8,-0.4) node[below] {$H$};
        \end{tikzpicture}
   \caption{Loop diagram that give rise to dimension-6 neutrino mass through Wilson coefficient $C_{\nu H}$.\label{fig:dim6triplettopology} }
\end{figure}
In this setup, there is a direct contribution to the dimension-4 mass at the loop level as well as to the dimension-6 mass at the tree level as shown in Fig.~\ref{fig:dim4tripletlooptopology}. However, these graphs can be easily suppressed by taking the limit $\mu \to 0$. In this limit, the contribution to the magnetic moment in Fig.~\ref{fig:topology}\,(a) is unchanged, while neutrino gets its mass induced from another dimension-6 topology at loop-level, as shown in Fig.~\ref{fig:dim6triplettopology}. As can be seen from the graph, this mass contribution involves the same couplings as the dipole term, with two additional powers of $y_T$, giving us a handle to separate the two contributions. By choosing $y_T$ to be small, the mass contribution can be rendered negligible, while still allowing a sizable magnetic moment.
    
\bibliographystyle{utcaps_mod}
\bibliography{references}
\end{document}